\documentclass[a4paper,10pt,twocolumn]{article}
\usepackage[utf8]{inputenc}
 \usepackage{graphicx}
\usepackage{color}
 \usepackage{epsfig}

\usepackage{amssymb}
\usepackage{amsmath}

\title{Pseudogap and the specific heat of high $T_c$ superconductors}
\author{E. J. Calegari\footnote{eleonir@ufsm.br}\\
Laborat\'orio de Teoria da Mat\'eria Condensada,\\
Departamento de F\'{\i}sica - UFSM, 97105-900, Santa Maria, RS, Brazil\\ \\
S. G. Magalhaes\\Instituto de F\'isica, Universidade Federal Fluminense\\ Av. Litor\^anea s/n, 24210,
346, Niter\'oi, Rio de Janeiro, Brazil\\ \\
C. M. Chaves and A. Troper\\Centro Brasileiro de Pesquisas F\'{\i}sicas\\ Rua Xavier Sigaud 150, 22290-180,
 Rio de Janeiro, RJ, Brazil
}

\begin{document}

\maketitle

\begin{abstract}
The specific heat of a two dimensional repulsive Hubbard model with local interaction is investigated. We use the two-pole 
approximation which exhibits explicitly important correlations that are sources of the pseudogap anomaly. The interplay between the 
specific heat and the pseudogap is the main focus of the present work. Our self consistent numerical results show that above the occupation $n_T\approx 0.85$, the specific heat starts to decrease due to the presence of a pseudogap in the density of states. We  have also observed a two peak structure in the specific heat. Such structure is robust with respect to the Coulomb interaction $U$ but it is significantly affected by the occupation $n_T$. 
A detailed study of the two peak structure is carried out in terms of the renormalized quasi-particle bands. The role of the
second nearest neighbor hopping on the specific heat behavior and on the pseudogap, is extensively discussed. 

\end{abstract}

\section{Introduction}
The one-band Hubbard model with repulsive local interaction \cite{Hubbard} is one of the simplest but significant lattice models 
which allows to describe a rich phenomenology of strong correlated electron systems. Such model has become even more important
after the discovery of the high temperature superconductors (HTSC) \cite{bednorz}. Nevertheless, although the two dimensional one-band Hubbard
model has been intensively investigated within different levels of approximation, there are still open issues, for instance, the existence of 
superconductivity and pseudogap  \cite{Timusk,cuprates,cuprates1}.

The nature of the pseudogap is one important key to understanding HTSC.
A substantial indication of pseudogap can be confirmed through the analysis of the specific heat behavior \cite{loram,rice}.
From the theoretical point of view, different approaches \cite{moreo,Bonca,avela} have been used to investigate the specific heat of the one-band Hubbard model. However, the focus
has been on the observed two peak structure rather than its connection with the pseudogap. Monte Carlo simulation \cite{TPaiva} and 
dynamical mean field approximation \cite{kusunose}  on the two dimensional one-band Hubbard Model showed evidences 
of a pseudogap. But such calculations were carried out only at half filling. 

 
In this work, we calculate the electronic specific heat out of the half-filling ($n_T<1$, where $n_T=n_{\sigma}+n_{-\sigma}$) 
for the local repulsive Hubbard model. 
The Green's functions have been obtained in a two-pole approximation \cite{Roth,Edwards} which takes into account essential correlations that allow to capture important properties of the pseudogap. In addition to the nearest neighbor hopping, the Hubbard model considered in the present work includes
hopping to second nearest neighbors. Such hopping enhances the spin-spin correlations that affect both the pseudogap and the specific heat structure. 
We present results for the normal phase, i. e., $T>T_c$ where $T_c$ is the superconducting critical temperature.
Firstly, the two peak structure of the specific heat is analyzed in terms of the renormalized quasiparticle band. Then, the pseudogap and its interplay
with the specific heat, is discussed.

\section{Model and general formulation}
 
The model investigated is the two-dimensional one-band Hubbard model \cite{Hubbard}
\begin{equation}
H=\sum_{\langle \langle ij \rangle\rangle \sigma} t_{ij}d_{i\sigma}^{\dag}d_{j\sigma} + \frac{U}{2}\sum_{i \sigma} n_{i,\sigma} n_{i,-\sigma}-\mu\sum_{i\sigma}n_{i\sigma}
\label{eqH1}
\end{equation}
where $d_{i\sigma }^{\dag }(d_{i\sigma })$ is the fermionic creation
(annihilation) operator at site $i$ with spin $\sigma
=\{\uparrow ,\downarrow \}$ and $n_{i,\sigma }=d_{i\sigma }^{\dag
}d_{i\sigma }$ is the number operator. The quantity $t_{ij}$ represents the hopping between sites $i$ and $j$
and $\langle \langle ...\rangle \rangle $ indicates the sum over the first
and second-nearest-neighbors of $i$ and $\mu $ is the chemical
potential.
$U$ is the repulsive Coulomb potential between the $d$ electrons localized at the same site $i$.
The bare dispersion relation is  
\begin{equation}
{\varepsilon }_{\vec{k}}=2t[\cos (k_{x}a) +\cos (
k_{y}a)] +4t_{2}\cos ( k_{x}a) \cos (k_{y}a) 
\end{equation}
where $t$ is the first-neighbor and $t_{2}$ is the second-neighbor hopping
amplitudes and $a$ is the lattice parameter.

In the two-pole approximation proposed by Roth \cite{Roth,Edwards}, the Green's function matrix is 
%
$\mathbf{G}\left( \omega \right) =\mathbf{N}\left( \omega \mathbf{N-E}\right)
^{-1}\mathbf{N} $
%
in which $\mathbf{N}$ and $\mathbf{E}$ are the normalization and the energy
matrices, respectively \cite{Edwards}.

Following the formalism discussed in reference \cite{Kishore}, 
the energy $E=\frac{\langle H\rangle}{N}$ can be written as ($N$ being the number of sites of the system):
%
\begin{align}
E&=\frac{i}{2N}\lim_{\delta \rightarrow
0^{+}}\sum_{{\vec{k}},\sigma }\int_{-\infty }^{\infty }f(\omega)\ ({
\omega }+{\mu }+{\varepsilon }_{\vec{k}})\nonumber\\
&\times[G_{{{\vec{k}},\sigma }}({\omega }+i{
\delta })-G_{{{\vec{k}},\sigma }}({\omega }-i{\delta })]d{\omega }
\label{eqE}
\end{align}
%
where $f(\omega)$ is the Fermi function and
the Green's function is
\begin{equation}
G_{{{\vec{k}},\sigma }}({\omega })=
%
\frac{Z_{1,\sigma }({{\vec{k}}})}{\omega
-\omega _{1,\sigma\vec{k}}}+\frac{Z_{2,\sigma }(\vec{k})}{\omega -\omega
_{2,\sigma\vec{k}}} 
\label{eqG}
\end{equation}
with 
%
\begin{equation}
 Z_{i,\sigma }(\vec{k})=\frac{1}{2}-(-1)^i\left[\frac{U(1-2n_{-\sigma})-\varepsilon_{\vec{k}}+W_{\vec{k},\sigma}}{2X_{\vec{k},\sigma}}\right]
 \label{Zi}
\end{equation}
and 
%
%
the renormalized bands
%
\begin{equation}
 \omega_{i,\sigma \vec{k}}=\frac{U+\varepsilon_{\vec{k}}+W_{\vec{k},\sigma}-2\mu}{2}
 +(-1)^i\left(\frac{X_{\vec{k},\sigma}}{2}\right)
 \label{wi}
\end{equation} 
%
%
where
%
$X_{\vec{k},\sigma}= \sqrt{(U-\varepsilon_{\vec{k}}+W_{\vec{k},\sigma})^2+4n_{-\sigma}U(\varepsilon_{\vec{k}}-W_{\vec{k},\sigma})}$.
%
%
%
In the grand canonical ensemble,
the energy is a function of the chemical potential $E\equiv E\left(\mu (T)\right)$,
where $\mu$ changes with the temperature. Therefore, the calculation of $C(T)$ defined as $C(T)=\frac{\partial {E}}{\partial {T}}$, must be performed keeping $\langle n\rangle$ 
constant in the $T-\mu$ plane \cite{moreo}.  
Combining equations  (\ref{eqE}), (\ref{eqG}) and $C(T)=\frac{\partial {E}}{\partial {T}}$ we get $C(T)=\int_{-\infty }^{\infty }F(\omega)d{\omega }$ where
\begin{equation}
F(\omega)=f'(\omega )g(\omega )
\label{B33}
\end{equation}
%
%
in which $f'(\omega )=\frac{1}{\omega}\frac{\partial {f(\omega)}}{\partial {T}}$ and $f(\omega)$ is the Fermi function.
The function $g(\omega )$ is defined as:
\begin{equation}
g(\omega )=\frac{1}{2N}\sum_{i=1}^{2}\sum_{{\vec{k}},\sigma }\widetilde{Z}_{i,\sigma }(\vec{k})\delta (\omega -\omega
_{i,\sigma\vec{k}})  
\label{B34}
\end{equation}
where
\begin{equation}
\widetilde{Z}_{i,\sigma }(\vec{k})=\left(\omega _{i,\sigma\vec{k}}+{\mu }+{\varepsilon }_{\vec{k}}\right)
\omega_{i,\sigma\vec{k}} Z_{i,\sigma }(\vec{k}).
\label{Zt}
\end{equation}

\begin{figure}[t]
\centering
\includegraphics[angle=-90,width=5cm]{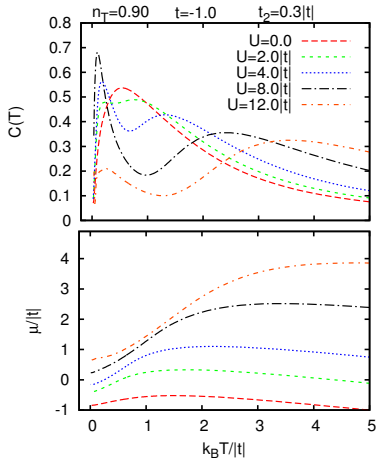}
\caption{In the upper panel, the specific heat as function of temperature for different values of $U$.The lower panel shows the chemical potential.}
\label{fig1}
\end{figure}

The band shift $W_{\vec{k}\sigma }$ introduced in equation (\ref{wi}), is:
%
$W_{\vec{k}\sigma } =\frac{1}{n_{\sigma}(1-n_{\sigma})}\frac{1}{N}\sum_{\vec{q}}\epsilon(\vec{k}-\vec{q})
F_{\sigma}(\vec{q})$,
%
where $\epsilon(\vec{k}-\vec{q})=\sum_{\langle\langle i=0\rangle\rangle j\neq 0}t_{0j}e^{i(\vec{k}-\vec{q})\cdot \vec{R}_j}$
and $F_{\sigma}(\vec{q})$ is given in terms of $\langle \vec{S_j}\cdot\vec{S_i}\rangle$, $n_{0j\sigma}=
\frac{1}{N}\sum_{\vec k}{\cal{F}}_{\omega}G_{\vec{k}\sigma}^{(11)}e^{i\vec{k}\cdot \vec{R}_j}$ and $m_{0j\sigma}=
\frac{1}{N}\sum_{\vec k}{\cal{F}}_{\omega}G_{\vec{k}\sigma}^{(12)}e^{i\vec{k}\cdot \vec{R}_j}$, 
where ${\cal{F}}_{\omega}\Gamma(\omega)\equiv\frac{1}{2\pi i}\oint d\omega f(\omega)\Gamma(\omega)$, in which $f(\omega)$ is the Fermi function 
and $\Gamma(\omega)$ a general Green's function. The Green's functions $G_{\vec{k}\sigma}^{(nm)}$ are obtained as in reference \cite{Calegari1}.
The spin-spin correlation function is given by:
\begin{equation}
\langle \vec{S_j}\cdot\vec{S_i}\rangle= h_{ij,-\sigma}-\frac{a_{ii\sigma}n_{-\sigma}}{2}
-\frac{a_{ij,-\sigma}n_{ij,\sigma}+b_{ij,-\sigma}m_{ij,\sigma}}{1+b_{ii,-\sigma}} \nonumber
\label{SjSi}
\end{equation}
%
where
%
$h_{ij,-\sigma}= \alpha_{\sigma}[\overline{n}-(a_{ij,-\sigma}n_{ij,-\sigma}+b_{ij,-\sigma}m_{ij,-\sigma})]$ 
%
%
%
with $\alpha_{\sigma}=(1-b_{ii\sigma})/[2(1-b_{ii-\sigma}b_{ii\sigma})]$ and  $\overline{n}=n_{-\sigma}^2(1-b_{ii-\sigma}b_{ii\sigma})$.
%
%
Also,
$a_{ij,-\sigma}=(n_{ij,-\sigma}-m_{ij,-\sigma})/(1-n_{\sigma})$ and $b_{ij,-\sigma}=(m_{ij,-\sigma}-n_{ij,-\sigma}n_{\sigma})/[n_{\sigma}(1-n_{\sigma})]$.
%
%

%
\begin{figure*}[t]
\centering
\includegraphics[angle=-90,width=9cm]{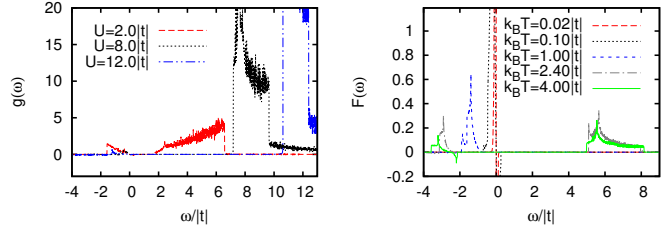}
\caption{In the left panel the behavior of $g(\omega)$ for $k_BT=0.03|t|$ and different values of $U$. The right panel shows
the function $F(\omega)$
for $U=8.0|t|$ and different temperatures. In both panels
$n_T=0.90$, $t=-1.0$ and $t_2=0.3|t|$.}
\label{gwU}
\end{figure*}

\section{Numerical results}
The upper panel in figure \ref{fig1} shows the specific heat as a function of temperature for $n_T=0.90$ and different values of $U$. For  $U=0.0$,
the specific heat presents a peak in the region of $k_BT\approx 0.55|t|$. This peak is known as the Schottky anomaly ( see ref. \cite{Xie} and references therein).
However, for $U=2.0|t|$ the Schottky anomaly starts to split giving rise to a two peak structure. So, for $U=4.0|t|$ the specific heat 
shows a peak at low temperature $k_BT\approx 0.2|t|$ and a second peak in $k_BT\approx 1.3|t|$. For $U>4.0|t|$, the specific heat 
has a two peak structure, i.e., the high temperature peak persist even in the strongly correlated limit.
\begin{figure*}[t]
\centering
\includegraphics[angle=-90,width=8.7cm]{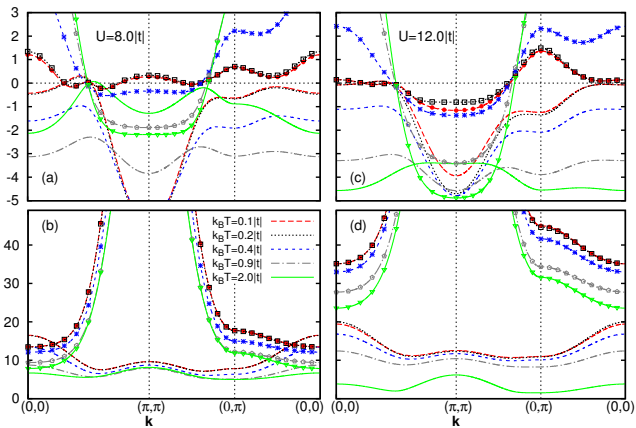}
\caption{ The lines with symbols show the effective spectral weight $\widetilde{Z}_{i,\sigma }(\vec{k})$ and the lines with no symbols show 
the renormalized bands $\omega_{i,\sigma }(\vec{k})$. The upper panels show $\widetilde{Z}_{1,\sigma }(\vec{k})$ and $\omega_{1,\sigma }(\vec{k})$ while 
the lower panels show $\widetilde{Z}_{2,\sigma }(\vec{k})$ and $\omega_{2,\sigma }(\vec{k})$. The panels (a) and (b) correspond to $U=8.0|t|$
and the panels (c) and (d) correspond to $U=12.0|t|$.}
\label{wZU}
\end{figure*}
The lower panel in figure \ref{fig1} shows the chemical potential $\mu$ as a function of temperature and different Coulomb interactions. The occupation is $n_T=0.90$, therefore, for 
$U=0$ the chemical potential is found below the van Hove singularity. As the temperature increases,  $\mu$ moves to higher energies in order to keep  
$n_T$ unchanged. However, when $\mu$ reaches the VHS (at $k_BT/|t|\backsimeq 1.1$), the high density of states in VHS forces the chemical potential back to 
lower energies to maintain $n_T$ unchanged. When $U\neq 0.0$, a gap opens on the density of states near the VHS and  $\mu$ increases with temperature
until the Fermi function reaches the upper Hubbard band. In this case, for a given temperature, $\mu$ can be found within the gap. 
If temperature increases more, the upper Hubbard band increases its contribution and  $\mu$ must return to lower energies to maintain $n_T$ unchanged.
The behavior of the chemical potential is closely related to the specific heat structure, mainly at low temperatures. We verify that the local minimum
in $C(T)$ occurs when $\mu$ is found within the gap due to $U$. 

We can understand the two peak structure on $C(T)$ 
if we 
analyze the functions $F(\omega)$, $f'(\omega)$ and $g(\omega)$ defined above.
The function $g(\omega)$ is shown in the left panel in figure \ref{gwU} for different Coulomb interactions. 
The function $g(\omega)$ is directly related to the renormalized  density of states, therefore presents a gap due to the Coulomb interaction $U$.
For the model parameters considered in figure \ref{gwU}, in general, $g(\omega)$ is related to the lower Hubbard band if $\omega \lesssim 0.50$
and to the upper Hubbard band if  $\omega \gtrsim 0.50$. 
We verified that at low temperature, the function $f'(\omega)$ is very concentrated on the chemical potential and vanishes quickly for energies that are not close to the 
chemical potential. Moreover, for $n_T<1$, the chemical potential is found to be localized in the lower Hubbard band. Then, at low temperatures only the lower Hubbard band contributes to the 
specific heat and therefore, the low temperature peak on the specific heat is associated to the lower Hubbard band. On the other hand, at high temperatures the function 
$f'(\omega)$ spreads out, reaches the upper Hubbard band and the specific heat presents a second peak due to the contribution of that band.

Furthermore, as can be verified in figure \ref{fig1}, the position of the second peak in the $k_BT$ axis 
moves to high temperatures as $U$ increases. This occurs because the upper Hubbard band moves to higher energies as $U$ increases. Consequently, $g(\omega)$ associated to the 
upper Hubbard band moves also to higher energies and $f'(\omega)$ can reach this region only if $k_BT$ is large enough. Indeed, we verified that the relation between 
the high temperature peak position and $U$ is $k_BT\approx \frac{U}{3}$.
%
%
%
The right panel in figure \ref{gwU} presents the function $F(\omega)$ (defined in equation (\ref{B33})) for different temperatures.
Due to the relation between $f'(\omega)$ and $F(\omega)$,
at low temperatures, $F(\omega)$ is concentrated around the chemical potential $\mu$ (localized in $\frac{\omega}{|t|}=0$). 
Moreover, $F(\omega)$ presents a negative region near $\mu$ that decreases with temperature resulting in an increasing of the specific heat.
As a consequence, a peak appears on the specific heat at low temperature.
For $k_BT=1.00|t|$, the negative region in $F(\omega)$ vanishes and the positive region decreases due to the behavior of $f'(\omega)$. 
This feature of $F(\omega)$ produces a small value for the specific heat at $k_BT=1.00|t|$.
Increasing the temperature, for instance $k_BT=2.40|t|$, 
the intensity of $F(\omega)$ is low for $\frac{\omega}{|t|}<0$, and a significant region of positive $F(\omega)$ emerges 
for  $\frac{\omega}{|t|}\gtrsim 5.0$. This occurs because
at high temperatures $f'(\omega)$ spreads out and reaches the region of $g(\omega)$ associated to the upper Hubbard band. This behavior of $F(\omega)$ 
enhances again the specific heat giving rise to the high temperature peak. For $k_BT=4.0|t|$, $F(\omega)$ acts as for $k_BT=2.40|t|$,
however, it decreases the intensity due to the decreasing of the $f'(\omega)$ intensity.  

%
%
%
\begin{figure}[t]
\centering
\includegraphics[angle=-90,width=6cm]{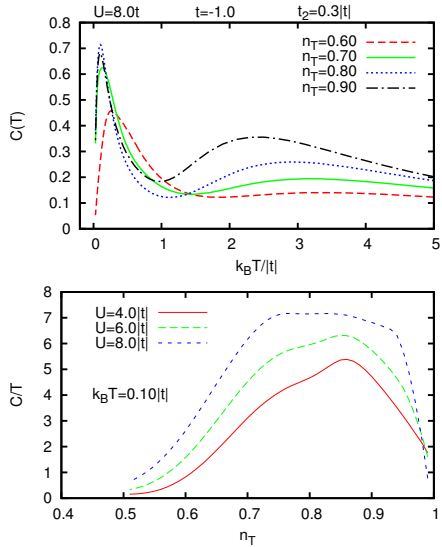}
\caption{In the upper panel, the specific heat for $U=8.0|t|$ and different occupations $n_T$. The lower panel shows the behavior of $\frac{C}{T}$ as a function of the occupation $n_T$.}
\label{CT}
\end{figure}

Figure \ref{wZU} shows the renormalized bands  $\omega_{i,\sigma\vec{k} }$ and 
the effective 
spectral weights $\widetilde{Z}_{i,\sigma }(\vec{k})$
defined in equations (\ref{wi}) and (\ref{Zt}).
At low temperatures
$\widetilde{Z}_{1,\sigma }(\vec{k})$ is negative only in a small region next to $(\frac{\pi}{2},\frac{\pi}{2})$. 
As temperature increases, this negative 
region enlarges along the direction $(\frac{\pi}{2},\frac{\pi}{2})$ - 
$(0,\pi)$. However, the intensity of $\widetilde{Z}_{1,\sigma }(\vec{k})$ is very small. In \ref{wZU}(c), the negative regions
of $\widetilde{Z}_{1,\sigma }(\vec{k})$ at low temperatures are greater than in \ref{wZU}(a), but its intensity remains small
when compared with the intensity of the positive regions.

An important feature observed in $\omega_{1,\sigma\vec{k}}$ is the presence 
of a pseudogap near $(0,\pi)$. This pseudogap plays an important role relative to the low temperature peak observed on the specific heat.
Indeed, the presence of a pseudogap affects strongly $\widetilde{Z}_{1,\sigma }(\vec{k})$ which is strongly dependent on  $\omega_{1,\sigma\vec{k}}$ 
(see equation (\ref{Zt}) ).  The opening of the pseudogap in $(0,\pi)$ moves the renormalized band $\omega_{1,\sigma\vec{k}}$ to negative energies.
As a consequence, $\widetilde{Z}_{1,\sigma }(\vec{k})$ is shifted to the positive region as can be observed, for instance, in figure \ref{wZU}(c), when  $k_BT=0.1|t|$.
In (b) and (d), $\widetilde{Z}_{2,\sigma }(\vec{k})$ is always positive and increases its
intensity with $U$. The pseudogap emerges when the correlations of antiferromagnetic character associated to the correlation 
function $\langle\vec{S}_i\cdot\vec{S}_j\rangle$ present in the band shift, 
become strong sufficiently to 
push down the renormalized quasi-particle band $\omega_{1,\sigma\vec{k}}$ in $(\pi,\pi)$ \cite{Calegari1,Korshunov}.
%
\begin{figure}[h]
\centering
\includegraphics[angle=-90,width=6cm]{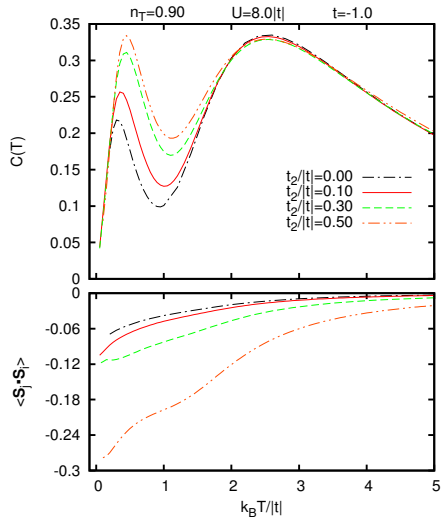}
\caption{ The upper panel shows the specific heat as a function of temperature for different values of $\frac{t_2}{|t|}$. The lower panel
shows the spin-spin correlations $\langle\vec{S}_i\cdot\vec{S}_j\rangle$ for the same parameters as in the upper panel.}
\label{fig0}
\end{figure} 
\begin{figure}[t]
\centering
\includegraphics[angle=-90,width=6cm]{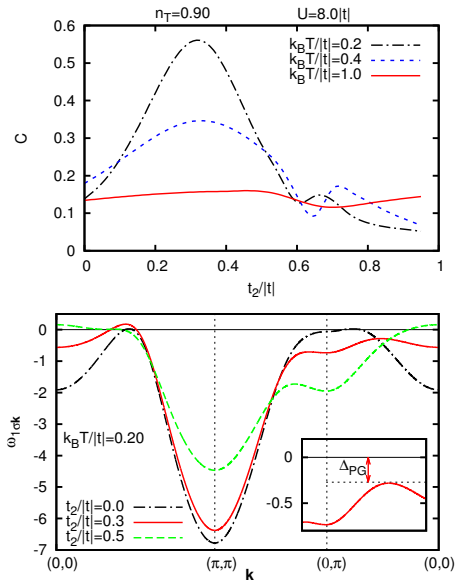}
\caption{The upper panel shows the specific heat has a function of $\frac{t_2}{|t|}$ and different temperatures.
The lower panel shows the renormalized quasi-particle band $\omega_{1\sigma\vec{k}}$ for $k_BT/|t|=0.20$ and the same parameters as in the upper panel. The inset
shows explicitly the pseudogap $\Delta_{PG}$.}
\label{Ct2}
\end{figure}
%

The upper panel in figure \ref{CT} shows the specific heat $C(T)$ as a function of temperature for different occupations. 
Notice that at low occupations $C(T)$ is characterized by a peak at low temperature. On the other hand, when  $n_T$
increases a second peak appears at high temperatures. 
At low $n_T$ only few electrons 
reach the upper Hubbard band and therefore the high temperature peak on $C(T)$ is negligible.
The behavior of $\frac{C}{T}$ as a function of $n_T$ is shown in the lower panel of figure \ref{CT}. This results agree qualitatively with those in reference \cite{markiewicz}.
We verified that for $n_T\gtrsim 0.85$ a pseudogap appears on the anti-nodal points of the renormalized quasi-particle bands (see the lower panel in figure \ref{Ct2}). 
Such pseudogap suppresses the density of states (DOS) on the chemical potential. As the specific heat $C(T)$ is directly related to the 
DOS, which in turn depends on the renormalized quasi-particle bands, the effects of the pseudogap appearing also on $C(T)$. 
In the present work, the function $F(\omega)$ defined in equation (\ref{B33})) associates 
the renormalized quasi-particle band and the specific heat. 
We verified that the pseudogap suppresses $F(\omega)$, below the chemical potential (in $\omega=0$), resulting in a decreasing in $C(T)$ for $n_T\gtrsim0.85$.
Therefore, it can be concluded that the decreasing of the specific heat above $n_T\simeq 0.85$ in figure  \ref{CT} is a clear manifestation of the pseudogap.

The effects of the second-nearest-neighbor $t_2$ on $C(T)$ are shown in the upper panel in figure  \ref{fig0}. We observe that the low temperature peak on  $C(T)$
is more affected by $t_2$ than the hight temperature peak. The lower panel shows that $t_2$
enhances significantly the spin-spin correlations $\langle\vec{S}_i\cdot\vec{S}_j\rangle$ which modifies the renormalized quasi-particle  bands
and consequently the function $F(\omega)$ and therefore, $C(T)$.  At low temperature $|\langle\vec{S}_i\cdot\vec{S}_j\rangle|$ is stronger and then its effects
on $C(T)$ are more evidenced than at high temperatures. This is the main reason why the low temperature peak on  $C(T)$ is more intensively affected by $t_2$.


The upper panel in figure \ref{Ct2} shows the specific heat as a function of the second nearest neighbor hopping amplitude $\frac{t_2}{|t|}$
at different temperatures. For $k_BT/|t|=0.20$ the specific heat presents a maximum for  $\frac{t_2}{|t|}\approx 0.30$ and then it decreases until
$\frac{t_2}{|t|}\simeq 0.60$. For  $\frac{t_2}{|t|}> 0.60$ the specific heat does not change significantly. The lower panel in figure \ref{Ct2} 
presents the renormalized quasi-particle  band $\omega_{1\sigma\vec{k}}$ for $k_BT/|t|=0.20$ and three distinct values of $\frac{t_2}{|t|}$. When $\frac{t_2}{|t|}= 0.30$,
a pseudogap is observed near the antinodal point $(0,\pi)$. If  $\frac{t_2}{|t|}$ increases the pseudogap closes as shown for $\frac{t_2}{|t|}= 0.50$.
There is a remarkable coincidence between the maximum on the specific heat and the maximum pseudogap. The analysis of $F(\omega)$ shows that 
it gives the greater contribution to $C(T)$ when $\frac{t_2}{|t|}= 0.30$. This occurs due to the wide flat region in $\omega_{1\sigma\vec{k}}$ which 
extends from $(\frac{\pi}{2},\pi)$ to approximately $(\frac{\pi}{4},\frac{\pi}{4})$ and gives rise to a very large $F(\omega)$ near
the chemical potential. Therefore, even that $t_2$ opens a pseudogap which suppresses  $F(\omega)$ on the chemical potential, it also
produces the flat region (near the chemical potential) that overcomes the effect of the pseudogap and increases significantly the specific heat.
At high temperatures the pseudogap closes and also the effect of the flat region is suppressed by temperature effects.    

\section{Conclusions}

In summary, the analysis of the two peak structure of $C(T)$ in terms of the renormalized quasi-particle bands allowed us  to investigate also a very important feature
present in the region of $n_T\gtrsim 0.85$, i. e. the pseudogap region. We observed that above $n_T\approx 0.85$, the specific
heat decreases signaling the pseudogap presence. In reference \cite{TPaiva}, the authors suggest the presence of a pseudogap at
half filling of the Hubbard model with Monte Carlo simulation. Here, we confirm the pseudogap existence 
also out of the half filling.

\subsection*{Acknowledgments}

This work was partially supported by the Brazilian agencies CNPq,
CAPES,
FAPERGS
and FAPERJ.


\end{document}